# Introduction to Laser Physics

*L. Corner*
Cockcroft Institute, University of Liverpool, UK

**Abstract**
This paper is a basic introduction to laser physics, especially that relevant to accelerator science. It presents the essential physics of a laser, some of the different types of laser system available, the propagation of laser beams, and the key diagnostics that are used for lasers.

## 1      Lasers

The acronym LASER stands for Light Amplification by Stimulated Emission of Radiation. It is beyond the scope of this paper to cover in depth the physics of laser operation; for more detailed information the reader is directed to the many excellent textbooks on the subject [1, 2, 3]. They are of interest because the light emitted from a laser has several distinctive features that separates it from that from other sources. For example, laser light is distinctly monochromatic compared to say, a lightbulb or the sun. It is directional, rather than being radiated isotropically, and importantly, spatially and temporally coherent to a much higher degree than other light sources.

The basic principle of lasing action is the stimulated transition of an electron in a high energy level to a lower one, emitting a photon with the same properties as the incident photon that initiated the transition. This requires that there are more electrons in the upper state than the lower, or the incident photon would be absorbed. This is called a 'population inversion' and is an unusual state, as a thermal Boltzmann distribution of population in an atom (where 'atom' can be taken to include ions and molecules) would be for there to be more electrons in the lower energy level.

Imagining the simplest case of a two level atom, we can see that it is impossible to establish a population inversion in this closed case. The rates of absorption and stimulated emission are proportional to the populations in the lower and upper levels respectively and to the strength of the applied field. In the steady state this means no net change in the population of either level and at most half the total population can be in the upper state. Therefore a population inversion cannot be created in a two level system and lasers need at least three participating energy levels.

### 1.1      Three and four level lasers

The three level laser is illustrated schematically in fig. 1. Here incident pump light is absorbed on the transition from the ground state level 0 → level 2. The population in level 2 relaxes into level 1 via a non-radiative transition. If more than half the population can be moved from the ground state by the pump, then it is possible to establish a population inversion between levels 1 and 0 which constitute the upper and lower laser levels. Clearly, it requires considerable pump energy to move enough population from the ground state to make a population inversion in a three level laser system and a more efficient arrangement is a four level laser also shown in fig. 1. Here, the pump transition is from level 0 (the ground state) to level 3, which then decays to level 2. If level 1 is sufficiently far above the ground state that it is not thermally populated then a single electron in level 2 constitutes a population inversion which clearly requires much less pump energy to establish than for a three level system. It is also advantageous for population in level 1 to rapidly decay back down to the ground state so that it remains effectively empty.

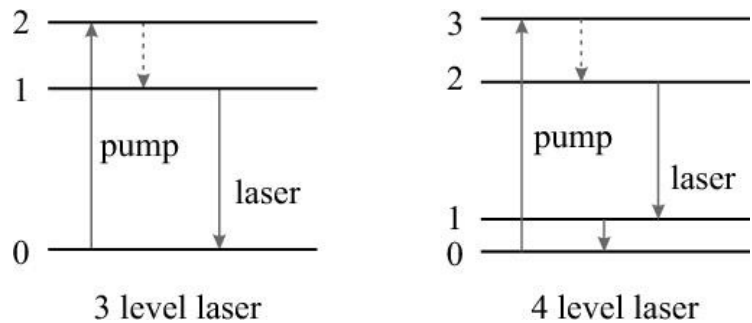

**Fig. 1**: Schematic illustration of three and four level lasers

These are of course simplified versions of real atomic or molecular systems, and not all laser media can be so neatly categorised; some media can be more three or four level like for different lasing wavelengths, and the lasing levels themselves are not infinitely sharp but are broadened in energy.

Depending on the exact properties of a particular laser medium, it may be possible to create a population inversion in the steady state and for the laser to operate continuous wave (cw). In some cases it is only possible to achieve population inversion temporarily and the laser emits pulses of light at a specific repetition rate. Some lasers are deliberately made to operate in pulsed mode to achieve higher peak powers than can be achieved using cw systems.

## 1.2 Broadening mechanisms

The transition on which a laser operates is by no means totally monochromatic, as implied by the narrow energy levels drawn in schematic representations of lasers such as fig. 1. In fact, one or both levels of a laser transition are subject to broadening by a number of different mechanisms and so the laser output will have a finite spectral width (its linewidth), with a characteristic spectral shape depending on the dominant broadening mechanism. These can be divided into two classes, homogeneous and inhomogeneous broadening. Homogeneous broadening includes the fundamental lifetime broadening which is a consequence of the finite lifetimes of the upper and lower lasing levels – essentially the uncertainty principle dictates that there must be an uncertainty in the energy of an excited state with a certain radiative lifetime and this causes the energy levels have a finite spectral width. So the lasing transition has an intrinsic linewidth that is the sum of the widths of the upper and lower levels, which is the fundamental minimum achievable and has a Lorentzian spectral distribution.

However, there are many additional broadening mechanisms that can act to increase the linewidth of a laser transition and it rare to find a laser linewidth as narrow as simply given by the lifetime broadening component. A major homogenous broadening mechanism is pressure broadening where collisions between the radiating atoms lead to changes in the spectral output of the emitted light and thus broadening of the emission across the macroscopic ensemble, which also gives a Lorentzian lineshape.

Some important broadening mechanisms are inhomogeneous i.e. they do not affect all atoms in the laser in the same way. One of these is Doppler broadening, which is important for gas lasers and is caused by the Doppler shift of the emitted laser light from the thermal motion of the atoms. Doppler broadening leads to a Gaussian output linewidth. Inhomogeneous broadening also often arises in solid state lasers, where the local environment of each lasing ion within a crystalline host or amorphous medium such as glass might be quite different, leading to significant spectral broadening of the laser output.

## 1.3 Laser oscillators

Establishing a population inversion in a laser medium is not enough to ensure lasing action. The basic form of a laser oscillator (a system that produces laser light with the properties outlined in section 1) consists of a laser gain medium, a pump source and a cavity, shown schematically in fig. 2. The pump source (which may be a flashlamp, electrical discharge, other laser etc.) creates a population inversion in the laser medium. Some of the upper state population decays via spontaneous emission and photons emitted close to the optical axis of the cavity are trapped between the mirrors and recirculate through the gain medium. These photons then stimulate further transitions from the upper state and the light field is amplified on every pass through the gain medium. Thus a large circulating body of photons builds up in the optical cavity. One of the cavity mirrors is partially transmissive, so some of the photons are emitted from this mirror (the 'output coupler') forming the laser beam. Once the round trip gain of the laser exceeds the cavity losses (through the output coupler, from absorption or diffraction loss) the laser is said to have reached threshold and it lases.

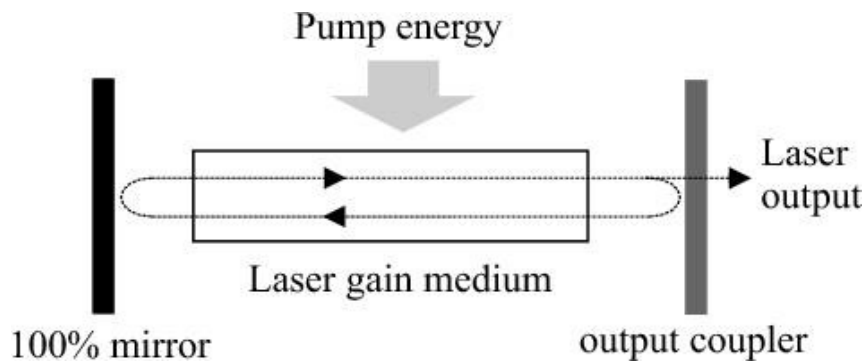

**Fig. 2:** Laser oscillator cavity

## 2 Types of laser

Lasers are normally categorised as gas, liquid or solid state, depending on the lasing medium. Of these, the most important are the solid state lasers, although there are significant systems in each of the other categories. In the following subsections, representative examples of each type of laser are described. More details about these lasers can be found in refs. [1,2,3,4].

### 2.1 Gas lasers

These use a gas as the laser medium, generally with the source of excitation (the pump) being an electrical discharge through the gas, creating ions and electrons. Some examples of gas lasers include:

#### 2.1.1 The helium-neon laser (HeNe)

This uses a mixture of helium and neon as the lasing medium and lases in the red at 632.8nm, although other wavelengths are possible. Helium-neon lasers are often found in optics labs for testing and alignment, owing to their relatively low cost and small footprint.

### 2.1.2 *The carbon dioxide ($CO_2$) laser*

These systems lase in the infrared at 10.6μm and have found particular use in industry for applications such as cutting and welding as they can achieve extremely high powers. $CO_2$ lasers are also used for some medical applications.

### 2.1.3 *Excimer lasers*

This is a general term for gas lasers where the gain medium is a mix of a noble gas, such as argon or krypton, and a reactive gas such as chlorine or fluorine. These lasers are useful because they lase in the ultraviolet (~126 – 350nm) where there are few other available laser sources. These lasers have important applications in photolithography and medicine.

## 2.2 Liquid lasers

These use a liquid as the lasing medium, usually an organic dye in a solvent. One of the most well known is rhodamine 6G, which can lase from ~560 – 630nm, producing light in the orange/yellow range of the visible spectrum where there are few other sources. Dye lasers have been important as the earliest tunable laser sources, which has been hugely influential in modern spectroscopy. The broad lasing bandwidth also allows dye lasers to support ultrashort pulsed operation. In general, although dye and other liquid lasers have been very important historically, the difficulty of handling a liquid gain medium means other laser sources are preferred if available.

## 2.3 Solid state lasers

This is perhaps the broadest category of laser, ranging from lasing ions doped into glass, crystal, or ceramic host materials with form factors ranging from building size large high power glass systems to optical fibre lasers and semiconductor diode systems.

### 2.3.1 *Semiconductor diode lasers*

These are perhaps the most important class of laser in terms of the numbers produced per year, being deployed in large numbers in CD and DVD players, optical telecommunications systems and barcode readers. The semiconductor lasing medium typically has dimensions of only a few tens of microns. The large reflection coefficient at the surfaces and high gain of these lasers mean they do not require external mirrors to create an oscillator cavity so they can be very small indeed, contributing to their widespread use. These lasers can operate on a variety of wavelengths across the visible and near-infrared spectrum depending on the exact semiconductor material used. The pump source for semiconductor lasers is electrical current, which means they have a high electrical to optical (wall plug) efficiency. Semiconductor diodes are important in high power laser science as pump sources for other lasers.

### 2.3.2 *Optical fibre lasers*

This term refers to the structure of the laser, rather than the specific gain medium. Optical fibres are waveguides, typically constructed of glass with a core surrounded by a cladding of lower refractive index, which can guide light over long distances. These passive fibres are well known for their use in high speed data communications but it is also possible to dope lasing ions into the core to create an active, i.e. amplifying, optical fibre laser. Some examples are erbium doped fibre amplifiers (EDFAs), which are vital component of long distance optical telecommunications, and ytterbium doped fibres which can operate at very high average powers with important applications in industry.

*2.3.3    Nd doped lasers*

Neodymium (Nd) is one of the most common lasing ions, and has been successfully used in a number of hosts, of which yttrium aluminium garnet (Nd:YAG) and glass are the best known. Nd:YAG is a widely used laser, operating at 1064nm and pumped using flashlamps or near infrared semiconductor laser diodes. It can operate in either cw or pulsed mode, typically producing pulses of tens of nanoseconds in Q-switched mode. Nd:YAG lasers have many applications in medicine and science but an important application for accelerator science is the use of the green frequency doubled output [21] at 532nm as pump light for titanium sapphire lasers.

*2.3.4    Titanium sapphire lasers (Ti:sapp)*

Ti:sapp is the major laser medium used for ultrashort pulsed systems. This is because Ti:sapp can lase across a wide spectral range (700 – 1000nm) and can therefore directly support pulses as short as a few femtoseconds. Ti:sapp lasers are typically centred around 800nm and pumped in the green, often by the frequency doubled output of a Nd:YAG laser, which may itself be pumped with flashlamps, or increasingly now by diode lasers. The ultrashort output pulses of Ti:sapp oscillator and amplifier systems means that they can achieve extremely high peak powers. Powers of hundreds of Terawatts are routinely available with commercial systems and the current world record (as of early 2020) for peak power of > 10PW was produced by a Ti:sapp laser installed at the Extreme Light Infrastructure Nuclear Physics (ELI-NP) facility in Romania.

## 2.4    Chirped Pulse Amplification

The key to achieving the large peak powers described above is the chirped pulse amplification (CPA) technique, invented by Donna Strickland and Gerard Mourou in 1985 [5] and for which they part-shared the Nobel prize in physics in 2018. CPA is a method which allows the amplification of short pulses to energies that would normally damage the laser gain medium by manipulating the temporal structure of the laser pulse.

A CPA laser system typically consists of a laser oscillator (the 'front end') which produces low energy, femtosecond duration pulses. These pulses are stretched in time to hundreds of picoseconds or even nanaoseconds using diffraction gratings, although other elements such as prisms or fibre stretchers can be used. The diffraction grating stretcher is set up such that the short wavelength components of the pulse travel a longer distance than the long wavelength components, so that at the output of the stretcher the envelope of the laser pulse has been broadened by several orders of magnitude and is chirped (different frequencies occur at different times under the laser envelope). The peak power of the pulse is thus significantly reduced and may be increased in energy, often in multiple further amplification steps, without damaging the gain medium. The energy output of these lasers can reach many Joules. The high energy, long pulses are compressed in duration using a grating pair with opposite dispersion to the stretcher, reducing the pulse duration to close to the original value. CPA lasers have been successful in producing extremely high peak powers and intensities. These high energy, short pulse lasers are widely used in accelerator research to drive high gradient laser plasma wakefield accelerators and have enabled the rapid progress in this research area, generating electron energies up to several GeV [6]. An excellent overview of high power CPA laser systems worldwide can be found in reference [7].

One current problem in using these high peak CPA laser systems for driving plasma accelerators is their low repetition rate. Systems may operate at only one shot every few minutes and even the state of the art is to have PW pulses at 1Hz or multi-TW systems operating at a few Hz. For many applications of plasma accelerators (medical or industrial imaging, particle colliders) it is clear that operation at multi-kHz rates would be preferable. Thus an active area of research in laser physics is the development of laser systems which are capable of producing both high peak (>TW) power but also high average power i.e. many thousands of pulses a second. The approaches to achieving high peak and average power lasers fall broadly into two categories. One is to improve the technology of 'single emitter' lasers, that is, systems where one laser beam is amplified in multiple stages in series. These technological developments include improvements to heat removal and thermal management of the laser amplifying medium, allowing them to be fired more often without thermal effects damaging the medium. Major developments are also taking place in pump technology, producing pump lasers that are higher power, more efficient, and can operate at higher repetition rates [8]. Research is also being carried out on the investigation of new laser ions and host materials to identify gain media that have better thermal or lasing properties than the current standard laser systems, whilst also having a broad enough bandwidth to support femtosecond pulses [9].

The other main area of research in high average and peak power lasers is the combination of multiple emitters into a single beam. This is particularly interesting for fibre laser amplifiers, which have excellent thermal handling capability, electrical to optical efficiency and can operate at high average powers in continuous wave operation. An individual fibre amplifier has a low (<mJ) pulse energy, but the combination of multiple pulsed emitters could allow the generation of high energy pulses at a high repetition rate, and some lasing ions doped in glass fibre e.g. Yb, can support < 200fs pulses, potentially leading to high peak powers. Numerous methods have been explored for multiple emitter amplification and combination, splitting low energy input pulses spatially [10,11] temporally [12,13,14] or spectrally [15] (or a combination of these [16]), amplifying them separately and then recombining them. Initial experiments are very promising, but coherent combination is still very much a research topic and has not yet demonstrated peak powers comparable to single emitter lasers.

In addition to research into the lasers themselves, the prospect of high repetition rate high peak power lasers require components capable of withstanding these high energies at kHz repetition rates. This is an especial issue for optics such as mirrors and gratings which are used to transport and compress the laser pulses which can be easily damaged at these intensities or deform due to the high thermal load. This affects the laser wavefront and the size and quality of the laser focus that can be achieved and thus limits the intensity that can be attained with these systems. Concomitantly, plasma sources for high repetition rate LWFA need to be developed that are not damaged within a few hundreds or thousands of laser shots and can be used for days or months at a time. In addition, on a higher control level the reliability and control of high power laser systems need to be improved so they can be operated as energy sources for particle accelerators on a consistent 24/7 basis. Detailed reports on the laser technology required for accelerators can be found in refs. [17,18].

## 3    Beam Physics

In this section the mode structure of lasers is described and equations presented for the propagation and focusing of laser beams.

### 3.1    Longitudinal modes

The optical cavity of a laser oscillator strongly influences the spatial and temporal properties of its output. The cavity is simply an optical resonator which supports certain longitudinal resonant modes, equally spaced in the frequency domain. The oscillator can only lase on the frequencies of these modes that are within the gain bandwidth of the laser. Whether the laser will operate on only one or many of these modes is largely determined by the predominant broadening mechanism of the particular laser medium. In general, homogeneously broadened lasers will operate on a single longitudinal mode (SLM) whereas inhomogeneously broadened systems can operate on multiple longitudinal modes, although these can be forced to run SLM by the introduction of additional losses for some cavity modes by, for example, using a Fabry-Pérot etalon in the cavity. Some lasers e.g. Ti:Sapp can operate on many thousands of cavity modes and these are very important for generating output trains of short pulses through the phenomenon of mode-locking.

Mode-locking refers to a situation where all the lasing modes of a cavity are forced to oscillate with the same phase, as opposed to with random phases. This locking of the modes together produces a single short pulse that circulates within the oscillator cavity. Each time the pulse reaches the output coupler a small proportion of the pulse is transmitted and so the output of the modelocked laser is a train of short pulses separated by the round trip time of the cavity. For a Ti:sapp oscillator the output pulses may be 30 – 40fs in duration with a repetition rate of tens of MHz and an output power of ~ 1W. These low energy, ultrashort pulsed modelocked oscillators typically form the front end of the CPA laser systems described in the previous section.

## 3.2   Transverse modes

The oscillator cavity also shapes the transverse spatial properties of the laser output. The transverse modes of the laser beam that are replicated after one round trip (an eigenmode of the cavity) are described by Laguerre-Gaussians for a system with cylindrical symmetry, and Hermite-Gaussians for systems with rectangular symmetry. These are in general more common owing to the breaking of cylindrical symmetry in the cavity by the introduction of optical components such as polarisers. In many cases it is desirable to restrict the output of the laser to the lowest order transverse mode which in both cases is a simple Gaussian. This is the most usual spatial output of a laser oscillator cavity and the reason why Gaussian beam optics is important in the study of lasers.

## 3.3   Gaussian beam propagation

A Gaussian laser beam can be described by the following equation:

$$E(x,y,z,t) = u(x,y,z)e^{j(kz-\omega t)} \tag{1}$$

Here $u$ is the transverse Gaussian profile which varies slowly with longitudinal position ($z$) and is given by:

$$u(x,y,z) = \frac{1}{q(z)}\exp\left[-jk\frac{x^2+y^2}{2}\frac{1}{q(z)}\right] \tag{2}$$

The parameter $q$ in this expression is called the 'complex beam parameter' and is determined by the beam size $w$ and the radius of curvature of the beam $R$:

$$\frac{1}{q(z)} = \frac{1}{R(z)} - j\frac{\lambda}{\pi w^2(z)} \tag{3}$$

These quantities are illustrated in fig. 3. It is important to note that the definition of the spot size $w$ is given by the $1/e^2$ radius of the intensity profile, not the diameter, and that this definition of the size of the laser beam will in general be different to the definitions of the size of charged particle beams used by accelerator scientists.

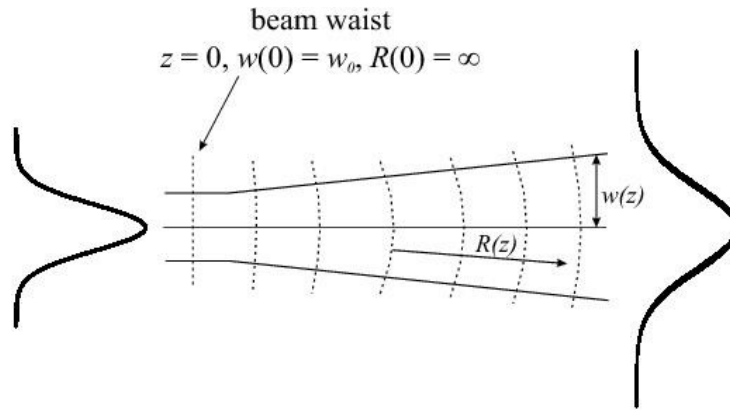

**Fig.** 3 Gaussian beam propagation

The change of spot size $w$ and radius of curvature $R$ with propagation distance $z$ from a focus are given by:

$$w(z) = w_0 \sqrt{1 + \left(\frac{z}{z_R}\right)^2} \qquad (4)$$

$$R(z) = z\left[1 + \left(\frac{z_R}{z}\right)^2\right] \qquad (5)$$

Here $w_o$ is the beam waist size at focus, and $z_R$ is the Rayleigh range given by:

$$z_R = \frac{\pi w_0^2}{\lambda} \qquad (6)$$

The Rayleigh range is an important parameter and is the distance over which the beam expands to double its initial area, shown in fig. 4. Twice the Rayleigh range, the confocal parameter $b$, is often used as an estimate of the distance over which a Gaussian beam is approximately collimated.

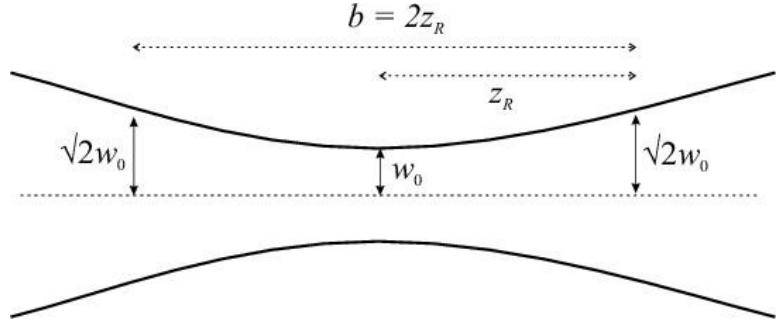

**Fig. 4:** Gaussian beam focal region

The equations above are for Gaussian beam propagation through free space. Ray transfer matrix analysis can be used to see how the beam transforms through an optical system [1]. This approach is illustrated in fig. 5.

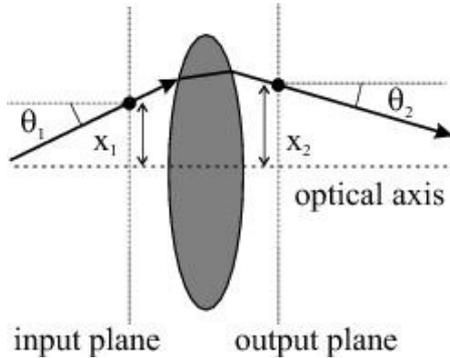

**Fig. 5**: Ray matrix analysis

We can define the position and angle, $x_1$, $\theta_1$, of an input light ray at the beginning of an optical system and find that it is related to the output position and angle $x_2$, $\theta_2$, by a 2 x 2 matrix describing the optical system as given below:

$$\begin{bmatrix} x_2 \\ \theta_2 \end{bmatrix} = \begin{bmatrix} A & B \\ C & D \end{bmatrix} \begin{bmatrix} x_1 \\ \theta_1 \end{bmatrix} \tag{7}$$

The ABCD matrix for a complex optical system can be found by multiplying together the matrices describing the component parts of the system, for example, free space propagation or transmission through a thin lens. This approach is derived from geometrical ray optics but can be usefully applied to Gaussian beam propagation by the following transformation:

$$q_2 = \frac{Aq_1 + B}{Cq_1 + D} \tag{8}$$

Thus if the $q$ parameter of a Gaussian beam and the relevant ABCD matrix are known it is possible to easily calculate the $q$ parameter, and hence the spot size and radius of curvature of the beam, elsewhere. This can be a powerful technique for designing optical systems to produce specific spot sizes for a given laser beam input.

A major requirement for lasers used in accelerator science is to produce a very small focused spot. This is important for generating high intensities for driving wakefields in a plasma, or for scanning over a charged particle beam to measure its size from the scattered photons. Achieving a small (<50 μm), high quality focus requires some care. An approximate formula for the focus of a Gaussian beam [3] is

$$w_0 = \frac{\lambda f}{\pi w_i} \tag{9}$$

where $f$ is the focal length of the optic and $w_i$ and $w_o$ the input and focussed spot sizes. This shows that to achieve a small spot size requires a short focal length, a short wavelength and a large input beam size. This ignores any effects of aberrations in the optical system or imperfections of the laser wavefront, and the actual spot size will usually be larger than predicted under the assumption of perfect optics.

For high power laser systems, the output beam is often a top hat spatial profile rather than a Gaussian. This allows more efficient extraction of gain from a cylindrical amplifier utilising the whole volume available. A common approach when discussing focusing high power beams is the concept of 'f number', written f/#, of a system. The f/# is given by the focal length of the optic (lens, parabola etc.) divided by its diameter, f/D. This is fixed for a specific optic but may not be the effective f/# of the optic in use, which is the focal length divided by the incident beam diameter. Thus a parabola with a focal length of 150mm and a diameter of 50mm has a f/# = 3 (150/50) but used with an input beam diameter of 30mm has an effective f/# = 5 (150/30). The focal spot size for such an optic can be estimated from $w \sim \lambda$ * effective f/# although again this does take into consideration the effect of any imperfections in the optical system or laser wavefront which often result in a much larger spot size (and thus lower intensity) than predicted.

It is possible to reduce these effects by the use of an adaptive optic. This is usually a reflective optic with a deformable surface which may be used to alter the laser wavefront to compensate for aberrations of the optical system and produce a smaller higher quality focus.

## 4    Diagnostics

A major part of any laser set up are the diagnostics. These are used both to ensure that the laser itself is working correctly, and are often also part of the data acquisition of the experiment. When planning an experiment it is wise to think about what diagnostics are required so they can be integrated into the set up from the beginning, and not inserted ad hoc later. Planning the diagnostics comes under two headings:

a) What needs to be measured? Examples could be wavelength, spatial mode quality, longitudinal mode, focused spot size, energy, power, pulse duration.

b) When and how often does the measurement need to be done? Once, daily, before and after a data set, every laser shot? In particular, if certain data (e.g. pulse duration, focus size, energy) are required for every laser shot the diagnostic has to be compatible with the main experiment and run at the same time. A measurement required only daily can be more intrusive if necessary.

## 4.1 Wavelength

This is normally measured with a spectrometer, of which there are many commercial options available. Spectrometers are nearly all based on using one or more diffraction gratings to disperse the incident light onto a detector. Some important considerations when choosing a spectrometer are its wavelength coverage and resolution. For simply checking that the laser is at approximately the correct wavelength, a smaller cheaper handheld device, often fibre coupled, might be appropriate. If more precise measurements of, for example, the red shift of a laser as it loses energy to a plasma, or of the spectral fringes from an interferometric diagnostic of the plasma density, are required, it is necessary to check that the grating and detector provide high enough resolution.

## 4.2 Power/Energy

Power (for cw lasers) or energy (pulsed lasers) meters are available from many vendors. For high repetition rate lasers, energy meters might not be able to measure each pulse individually so a power meter can be used to measure the average power output and then divide by the number of pulses to find the individual pulse energy.

A common approach is to place a meter behind a partially reflective optic (for example, a 99% reflective mirror). This enables the measurement of laser energy on shot while avoiding obstructing the experiment by blocking the laser beam or damaging the meter by exposing it to the full laser power.

## 4.3 Focus size/Spatial quality

It is useful to have cameras to measure the spatial quality of the laser focus and its jitter. Measurement of the laser beam position at several points in a beamline from laser to experiment can be used in a feedback loop with motorised optic mounts to stabilise the laser spot spatially. A camera, or a wavefront sensor, can also be used to monitor the quality of the laser spot at focus or indeed throughout the focal region and to provide feedback to an adaptive optic to maintain the required performance, which might be to maximise laser intensity, or maintain a specific focus shape. For small focal spot sizes magnification will be required to expand the image to a measurable size on a camera with finite pixel size.

## 4.4 Pulse duration

The measurement of laser pulse duration can be difficult. For pulse durations of nanoseconds it is possible to directly measure the temporal profile of a laser pulse using a fast photodiode and an oscilloscope, although it is worth noting that oscilloscopes with a high enough frequency response to measure a short laser pulse become extremely expensive, and that care should be taken to use the correct cables suitable for high frequency and correct impedance matching to make accurate measurements.

To measure ultrashort laser pulses different methods are required. Autocorrelation is one of the simplest of these [19,20]. This method uses a nonlinear optical crystal [4, 21] to gate the pulse as the second harmonic signal is only generated when the two copies of the laser pulse are coincident on the crystal at the same time. This signal can be measured with a slow photodetector and the duration of the autocorrelation signal deduced from the displacement of the moving arm in the setup. This width is then deconvoluted by a factor dependent on the assumed pulse shape to give the duration of the laser pulse. The autocorrelation method is limited by the fundamental assumption of the input laser pulse shape, and also the need to scan the moving arm, which means the autocorrelation signal is made up of the contributions of many laser pulses which may have slightly different shapes or jitter. It is possible to circumvent the last of these issues by using an arrangement to measure the autocorrelation of a laser pulse in a single shot [22,23].

A detailed discussion of the theory of laser pulse measurement can be found in refs. [24,25], but in summary to accurately reconstruct the electric field of a laser pulse that is too fast to measure directly and hence extract its duration it is necessary to measure both the spectral phase and amplitude. Autocorrelation does not retrieve the phase of the laser pulse, and so to obtain an accurate measurement of the duration of an ultrashort pulse other techniques are used. There are a number of refinements of these available but in particular commercially available devices can be categorised into either direct reconstruction techniques e.g. spectral phase interferometry for direct electric-field reconstruction (SPIDER) [26] or algorithmic methods such as frequency resolved optical gating (FROG) and its developments such as GRENOUILLE (grating-eliminated no-nonsense observation of ultrafast incident laser light e-fields) [27,28].

With all diagnostics it is important to make the measurement as close to the experiment, or under the same conditions as the experiment, as is possible. Even something as simple as propagation through a window can significantly distort the spatial or temporal properties of a laser beam. It is also important to ensure that all diagnostics are properly calibrated.

## 5  Conclusion

Lasers play an essential role in modern accelerator science, with applications ranging from charged particle beam diagnostics to drivers of novel high gradient plasma wakefield accelerators. This paper is a short introduction to the basics of laser physics, beam propagation and laser diagnostics, but cannot cover these topics in any depth. For more detail the reader is directed to the references.